\documentclass[a4paper,11pt]{article}
\usepackage{jcappub}
\usepackage{amsmath,amssymb,amsfonts,epsfig}
\usepackage{graphicx,epsfig,color}
\usepackage{subfigure}
\usepackage{natbib}
\usepackage{bm}

\newcommand{\be}{\begin{equation}}
\newcommand{\ee}{\end{equation}}

\def\prd{Phys.\ Rev.\ D}
\def\prl{Phys.\ Rev.\ Lett.}
\def\apj{Astrophys.\ J.}
\def\apjl{Astrophys.\ J.\ Lett.}
\def\apjs{Astrophys.\ J.\ Suppl.}
\def\mnras{Mon.\ Not.\ R.\ Astr.\ Soc.}
\def\aap{Astr.\ \& Astrophys.}

\def\jcap{JCAP}
\def\ijmpd{Int.\ J.\ Mod.\ Phys.\ D}

\newcommand{\Nside}{N_\text{side}}
\newcommand{\lmax}{{\ell_\text{max}}}
\newcommand{\uK}{~\mu\text{K}}
\newcommand{\LCDM}{$\Lambda$CDM}
\renewcommand{\deg}{^\circ}

\newcommand{\halfW}{0.475\textwidth}
\newcommand{\thirdW}{0.299\textwidth}

\begin{document}

\title{The Kullback--Leibler Divergence as an Estimator of the Statistical Properties of CMB Maps}

\author[a,b]{Assaf Ben-David,}
\author[b,c]{Hao Liu}
\author[a]{and Andrew D. Jackson}

\affiliation[a]{Niels Bohr International Academy, The Niels Bohr Institute, Blegdamsvej~17, DK-2100 Copenhagen~\O, Denmark}
\affiliation[b]{Discovery Center, The Niels Bohr Institute, Blegdamsvej~17, DK-2100 Copenhagen~\O, Denmark}
\affiliation[c]{Key Laboratory of Particle Astrophysics, Institute of High Energy Physics, CAS, China}

\emailAdd{bendavid@nbi.dk}
\emailAdd{liuhao@nbi.dk}

\abstract{
The identification of unsubtracted foreground residuals in the cosmic microwave background maps on large scales is of crucial importance for the analysis of polarization signals. These residuals add a non-Gaussian contribution to the data. We propose the Kullback--Leibler (KL) divergence as an effective, non-parametric test on the one-point probability distribution function of the data. With motivation in information theory, the KL divergence takes into account the entire range of the distribution and is highly non-local. We demonstrate its use by analyzing the large scales of the Planck 2013 SMICA temperature fluctuation map and find it consistent with the expected distribution at a level of 6\%. Comparing the results to those obtained using the more popular Kolmogorov--Smirnov test, we find the two methods to be in general agreement.
}


\maketitle
\flushbottom

\section{Introduction} % (fold)
\label{sec:introduction}

The Planck 2013 and 2015 data releases open new directions in precision cosmology with regard to a more advanced
investigation of the statistical isotropy and non-Gaussianity of the cosmic microwave background (CMB)~\cite{Planck1, Planck2, Planck_is, Planck_is2015}.
While generally confirming the Gaussianity and the statistical isotropy of the CMB (for the relevant multipole domain $\ell\ge 50$), the Planck
science team confirmed the existence of a variety of anomalies in the temperature anisotropy on large angular scales ($\ell\le 50$)
previously seen in WMAP data~\cite{Planck_is,Planck_is2015}.  Among these are the lack of power in the quadrupole component~\cite{low_quadrupole} (see, however,~\cite{wmap9b}
and~\cite{Planck_is}), the alignment of the quadrupole and octupole components~\cite{Copi1, wmap9b, Planck_is}, the unusual symmetry of the
octupole~\cite{mhansen}, anisotropies in the temperature angular power spectrum~\cite{WMAP7:powerspectra, Planck_is, Hansen2004, Eriksen2003},
 preferred directions~\cite{Pref.Direction1, Pref.Direction2, Copi2, Planck_is, Planck_is2015, Akrami2014}, asymmetry in the power of even and odd modes~\cite{parity1, parity2, wmap9b, Planck_is, Planck_is2015} and the Cold Spot~\cite{vielva2003, Coldspot1, Planck_is, Planck_is2015}.
 Some of these anomalies are probably a consequence of the residuals of foreground effects that could be a major source of contamination in the primordial $E$- and $B$-modes of polarization (in this connection see~\cite{Planck_bicep}).

The statistics of $B$-mode polarization that can be derived from ongoing and planned CMB experiments will be crucial for the determination of the
cosmological gravitational waves associated with inflation~\cite{inflation} at the range of multipoles $50\le \ell\le 150$, closer to
the domain of interest for BICEP2 and Planck~\cite{bicep,Planck_bicep}.   It seems likely that $B$-mode polarization in this range is affected
by Galactic dust emission, the statistical properties of which are very poorly known. (For $\ell >150$ we expect contamination of the $B$-modes
due to lensing effects the precise nature of which is also not fully understood.)  In the absence of such knowledge it is difficult to make \emph{a priori}
proposals for the best estimator of non-Gaussianity and statistical anisotropies in the derived $B$-modes due to possible contamination by foreground residuals.
Thus, we believe that it is of value to propose additional model-independent tests aimed at providing an improved quantitative understanding of the magnitude of non-Gaussianity in current CMB data.  Such tests would also be useful for the analysis of forthcoming CMB data sets.   In this paper
we propose use of the Kullback--Leibler (KL) divergence as such a test.  The goal of this paper is to illustrate the utility of the KL divergence in studying the properties of the CMB signal --- a Gaussian or almost Gaussian signal. The KL divergence is likely to be even more useful for very non-Gaussian cases such as the statistical behaviour of the Minkowski functionals for a single map or the pixel-pixel cross-correlation coefficient between two maps, when calculated in small areas.  We will consider such issues in a separate publication.

The one-point probability distribution function (PDF) would seem to be a reasonable starting point for the investigation of non-Gaussianity.
Such tests have been applied by the Planck team to a variety of derived CMB temperature maps including the SMICA, NILC, SEVEM and
Commander maps~\cite{Planck_is,Planck_is2015}.  In practice, these tests involve comparison of the various temperature fluctuation maps with an ensemble
of simulated maps.  The CMB temperature is characterized by a power spectrum, $C_\ell^\text{Planck}$, which
corresponds to the Planck 2013 concordance \LCDM\ model with the cosmological parameters listed in~\cite{Planck_cp}.
The simulated maps are obtained as Monte Carlo (MC) Gaussian draws on this power spectrum (in harmonic space).  When processed with the Planck component separation tool, the resulting simulation maps contain both CMB information as well as various residuals from foregrounds, the uncertainties of instrumental effects, etc.  For the Planck 2013 data release, the corresponding $10^3$ full focal plane simulations are referred to as the FFP7 simulations.
They reflect the intrinsic properties of the SMICA, NILC, SEVEM and Commander maps.  Differences between the FFP7 maps and the various empirical maps can provide useful information regarding non-Gaussianity.  In the following, we shall restrict our attention to the SMICA map.

In practice, the non-parametric Kolmogorov--Smirnov (KS) test is often used to assess the similarity of two distributions.  The KS test
characterizes the difference between the two cumulative distribution functions (CDF) in terms of the maximum absolute deviation between them.  The KS estimator, $\kappa$, is defined as
\begin{equation}
\kappa= \sqrt{n} \max[|F(x)-F_n(x)|] \, ,
\label{eq:KS_definition}
\end{equation}
where $F(x)$ is the theoretical expectation of the CDF and $F_n(x)$ is obtained from a data sample with $n$ elements.  Here, both $F(x)$ and $F_n(x)$ must  be normalized to the range $[0,1]$. Note that $F(x)$ is normally a continuous function or should at least be defined for all possible values of the data sample $F_n(x)$.   It is clear from Eq.~\eqref{eq:KS_definition} that the KS estimator $\kappa$ is local in the sense that its value will be determined at a point, $x$, where the PDFs corresponding to $F(x)$ and $F_n(x)$ cross.  We note that the use of PDFs in Eq.~\eqref{eq:KS_definition} instead of CDFs would result in a
maximal sensitivity to the largest local anomaly.  

Unlike the case of vectors (where the scalar product provides a standard measure), there is no generic ``best''
measure for quantifying the similarity of two distributions.  Thus, we believe that it is also useful to consider the Kullback--Leibler divergence for two discrete probability distributions, $P$ and $Q$.  The KL divergence on the set of points $i$ is defined~\cite{kl} as
\be \label{eq:KL_definition}
K(P\|Q)=\sum_i P_i\log\left(\frac{P_i}{Q_i}\right) \, .
\ee
In other words, the KL divergence is the expectation value of the logarithmic difference between the two probability distributions as computed with
weights of $P_i$.  Typically, $P$ represents the distribution of the data, while $Q$ represents a theoretical expectation of the data.  Unlike the
KS test, the KL divergence is non-local.  Indeed, we shall indicate below that it is in a sense ``maximally'' non-local.  It is familiar in information
theory, where it represents the difference between the intrinsic entropy,
$H_P$, of the distribution $P$ and the cross-entropy, $H_{PQ}$, between $P$ and $Q$,
\be
K(P\|Q)=H_{PQ}-H_P, \qquad H_P=-\sum_i P_i\log P_i, \qquad H_{PQ}=-\sum_i P_i\log Q_i \, .
\label{eq:entropy}
\ee
In more practical terms, consider the most probable result of $N$ independent random draws on the distribution $P$.  When $N$ is large, the number of
draws at point $i$ is simply $n_i = N P_i$.  Now construct the probabilities, $\Pi_P$ and $\Pi_Q$, that this most probable result was drawn at random
on distribution $P$ or $Q$, respectively.  The KL divergence of Eq.~\eqref{eq:KL_definition} is simply $N^{-1}\log{(\Pi_P / \Pi_Q)}$.
 We note that simulations of the CMB
map, drawn independently in harmonic space, have correlations in pixel space.
 Nevertheless we regard this argument as motivation for applying the KL divergence to CMB pixels.

The main goal of this paper is to illustrate the implementation of the KL divergence for
the analysis of the statistical properties of the derived CMB maps in the low multipole domain as a complementary test to the methods listed in~\cite{Planck_is,Planck_is2015}.  The structure of the paper is as follows. In Section~\ref{sec:KL_divergence_and_Planck_data} we present some properties of the KL divergence. We also use it to analyze the Planck SMICA map and compare it to both the FFP7 set and to a purely Gaussian ensemble. In addition, we compare the two ensembles and compare the KL divergence to the KS test in the low multipole domain of the CMB map. In Section~\ref{sec:discussion} we discuss the results. Note that the Planck papers~\cite{Planck_is,Planck_is2015} test the Gaussianity of the one-point PDF by analyzing its variance, skewness and kurtosis. In this sense, the KL divergence is simply another test on the global shape of the PDF.  Here, we restrict our analysis by using the SMICA map and the corresponding simulations.  The extension of the method to $E$- and $B$-modes of polarization does not require any modification.

\section{KL divergence and Planck data}
\label{sec:KL_divergence_and_Planck_data}

\subsection{Preliminary remarks: Properties of the KL divergence}
\label{subsec:Preliminary_remarks_KL_divergence}

\begin{figure}
	\centering
	\includegraphics[width=\halfW]{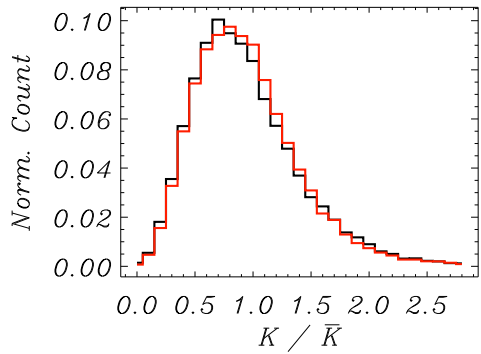}
	\caption{Histogram of KL divergences with $M=20000$ for $N=100$ (\emph{black})
	and $N=1000$ (\emph{red}).  Note that the horizontal axis is measured in units of the
	corresponding value of $\overline{K}$.}
	\label{fig:KL_hist_change_N}
\end{figure}

As noted above, the KL divergence provides a measure of the similarity of two known distributions.  In many cases, however, one of these
distributions is not known and must rather be approximated by the average PDF for a statistical ensemble of realizations of the random field.  The
question then arises of how closely the resulting proxy reflects the properties of the true underlying distribution.  To offer some insight in this
matter, we consider a toy model based on a discrete Gaussian distribution, $P_k$, with
\be
P_k \sim \exp{\left[ -  k^2/25 \right]}\, ,
\ee
and $k$ an integer between $-10$ and $+10$ subject to the obvious normalization condition.  The mean value of
$k$ is $0$ and the variance is approximately $25/2$.  Suppose for simplicity that we define an individual data
set as $N$ random draws on this distribution.  Each such data set can be regarded as a proxy, $Q_i$ for the underlying
distribution, and can be used to calculate the KL divergence $K(P\|Q)$ defined in Eq.~\eqref{eq:KL_definition}.  For a given value of $N$, we  repeat this
process $M$ times and compute the average KL divergence, $\overline{K}$, and the root-mean-square (RMS) deviation of the KL divergence, $\Delta K = \sqrt{\overline{K^2} - \overline{K}{}^2 }$.  The results of
this exercise are shown in Table~\ref{tab:Change_N}, where we have used the common value of $M=1000$.
\begin{table}
	\centering
	\caption{Mean and RMS values of the KL divergence for a discrete Gaussian distribution.}
	\vspace{10pt}
	\begin{tabular}{rcccc}
		\hline
		\noalign{\smallskip}
		\multicolumn{1}{c}{$N$} & $\overline{K}$ & $\Delta K$ & $N\overline{K}$ & $N\Delta K$ \\
		\hline
		 4000 & 0.002540 & 0.000796 & 10.160 & 3.18 \\
		 8000 & 0.001253 & 0.000413 & 10.024 & 3.30 \\
		16000 & 0.000636 & 0.000200 & 10.176 & 3.20 \\
		\hline
	\end{tabular}
	\label{tab:Change_N}
\end{table}
Several things seem clear.  Both the average value of the KL divergence and the RMS deviation from this average value vanish like $1/N$ for large $N$. From general arguments,
the KL divergence cannot be negative.  Fig.~\ref{fig:KL_hist_change_N} shows a histogram of the distribution of  KL divergences
obtained with $M = 20000$ for the cases $N=100$ and $N=1000$.
The KL divergences here are measured
in units of the corresponding value of $\overline{K}$.  The fact that these distributions scale like $1/N$ is
obvious.  These histograms suggest power law suppression near zero and Gaussian behaviour
for large values of the KL divergence.

The results shown in Fig.~\ref{fig:KL_hist_change_N} are not specific to the KL divergence, and qualitatively
similar results would be obtained for any measure chosen to describe the similarity of two distributions.
All that is required is that the measure chosen is always positive and vanishes when the distributions
being compared are identical.  When drawn as here, each individual data set can be thought of
as a combination of the ``exact'' distribution plus the amplitudes of $(N-1)$ ``fluctuations''.\footnote{Due to
the fact that each data set contains exactly $N$ draws.}  Obviously, the amplitudes of every one of these
fluctuations must be exactly zero if the measure is to be $K = 0$.  Of course, there are many combinations of the
fluctuation amplitudes that will give any fixed non-zero value of the measure, and their number increases as
$K$ grows from zero.  In contrast, for the case of only two options (e.g., ``heads'' and ``tails'') subject to a constraint
on the total number of draws, there is only a single degree of freedom.  In this case $K = 0$ is actually the most
probable value.

A few additional remarks can help clarify the properties of the KL divergence.  Consider that each individual
data set is sufficiently large that $Q_i = P_i + \delta P_i$ where $\delta P_i$ is small.  Under these conditions,
terms linear in $\delta P_i$ vanish as a consequence of normalization, and the KL divergence is given simply as
\be
K = \frac{1}{2} \, \sum_i \, \frac{\delta P_i^2}{P_i} \, .
\label{eq:KL_small_fluctuation}
\ee
We see that the distributions $P$ and $Q$ are now treated symmetrically in spite of the asymmetry that is
apparent in Eq.~\eqref{eq:KL_definition}.  Elementary arguments suggest that the average number of draws on bin $i$ will be
$N P_i \pm \sqrt{N P_i}$.   The corresponding proxy for the underlying distribution will be $P_i \pm \sqrt{P_i / N}$.
Given this result, Eq.~\eqref{eq:KL_small_fluctuation} suggests that, for fixed bin sizes and in the limit $N \to \infty$, each bin will make
a contribution to the KL divergence of roughly equal size.  It is hard to imagine a greater degree of non-locality. Moreover, in this limit the KL divergence is expected to be of order $N_b/N$, where $N_b$ is the number of bins.  In other words, if we define
\be \label{eq:alpha_definition}
\alpha=\frac{N}{N_b}K\, ,
\ee
we expect that $\alpha\sim\mathcal{O}(1)$. This realization allows us to assign a rough scale for the expected KL divergence. If two given distributions yield a much larger value of $\alpha$, we can conclude that they differ significantly from each other without resorting to comparisons with an ensemble. In the example shown in Table~\ref{tab:Change_N} we are using $N_b=30$, meaning $\alpha$ is small. This is to be expected since we are indeed using the true distribution to draw the data under examination.

\subsection{Preliminary remarks: Planck data}
\label{subsec:Preliminary_remarks_Planck_data}

We have performed the KL divergence test on the CMB map obtained  by the Planck collaboration~\cite{Planck_cp} using the SMICA method. Since we are interested in the
statistical properties of the CMB map on large scales, we first degrade the map from its native \textsc{HEALPix}~\cite{HEALPix} resolution of $\Nside=2048$
to $\Nside=32$. We then construct the convolution of this map with a Gaussian smoothing kernel of $5\deg$ FWHM and retain only the harmonic coefficients with
$\ell \le \lmax=96$.  The SMICA map provides a useful estimation of the CMB temperature fluctuations for a very large fraction --- but not all --- of the sky. We use the SMICA inpainting mask to exclude heavily contaminated regions, mainly the Galactic plane. At the resolution considered here, the mask removes about 6\% of the pixels, leaving the number of pixels under consideration to be $N=11565$. Since the analysis is performed in pixel space,
application of the mask is trivial.

In order to estimate the statistical significance of our results, we compare them to ensembles of realizations. In this work we use two different ensembles.
This is done in order to cross-check the significance estimations and also allows us to compare the two ensembles. The first of these ensembles is the
FFP7 set described above.  We degrade and smooth the FFP7 maps in the same manner as we did the SMICA map.  We expect that the effects of detector noise will be minor on the large scales considered here.  To test this expectation we therefore also make use of the best-fit power spectrum,
$C_\ell^\text{Planck}$~\cite{Planck_cp}, to generate an ensemble of $10^3$ Gaussian random realizations free of residuals.  As in the
case of the FFP7 ensemble, we restrict the multipole domain to $\lmax=96$ and smooth the harmonic coefficients with a Gaussian filter of $5\deg$ FWHM.
We also multiply the coefficients with the pixel window function associated with an $\Nside=32$ pixelization before converting them to
an $\Nside=32$ map.

\begin{figure}
	\centering
	\subfigure[]{
		\includegraphics[width=\halfW]{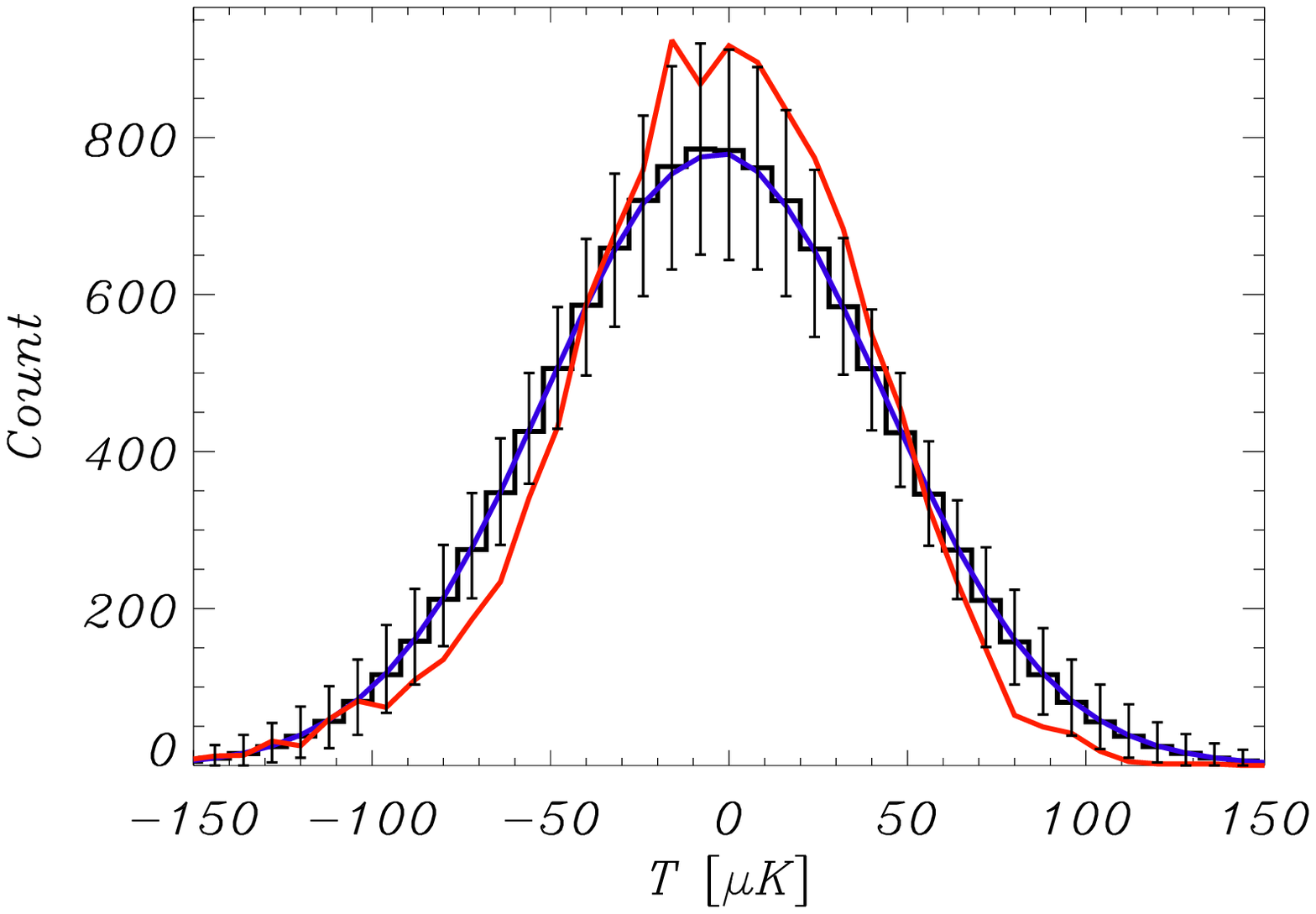}
		\label{subfig:maps_histograms}
	}
	\subfigure[]{
		\includegraphics[width=\halfW]{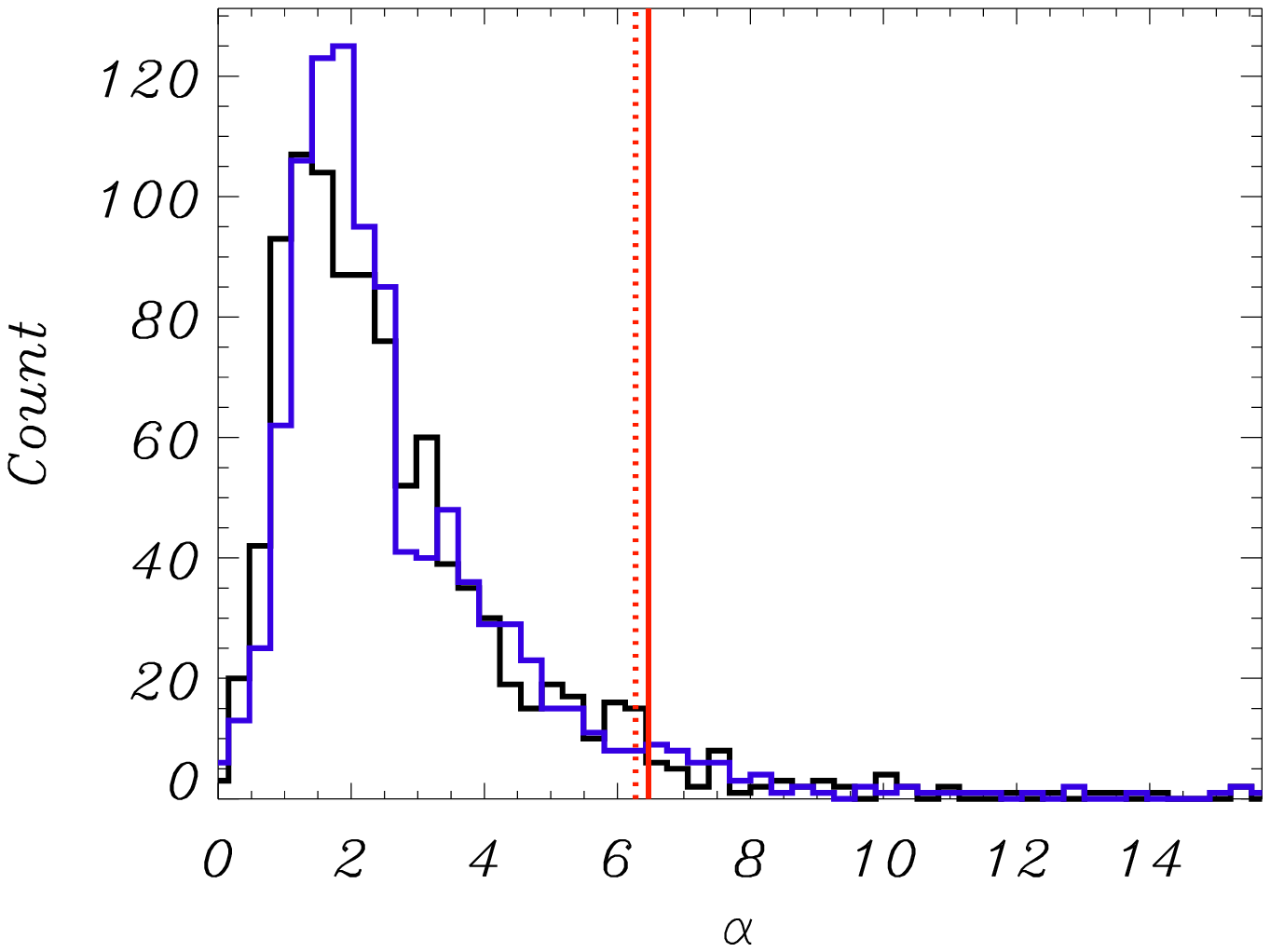}
		\label{subfig:KL_histograms}}
	 \caption{\subref{subfig:maps_histograms}~Number of counts versus amplitude of the SMICA map (\emph{red}), the FFP7 (\emph{black}) and
	the Gaussian (\emph{blue}) ensembles. The error bars show the 5--95\% range for the FFP7 set. \subref{subfig:KL_histograms}~Histograms of normalized KL
	divergence values for the FFP7 ensemble (\emph{black}) and the Gaussian ensemble (\emph{blue}). The values for the SMICA map are shown as red vertical
	lines, compared to the FFP7 mean distribution (\emph{solid}) and to the Gaussian mean distribution (\emph{dotted}).}
	\label{fig:KLResults}
\end{figure}

In our analysis we calculate the KL divergence for the SMICA map in pixel space. We calculate a histogram of the temperature fluctuations in the
unmasked pixels by taking bins of width $8\uK$ in the range $[-200,200]\uK$, meaning that the number of bins is $N_b=51$. Values outside this range are attributed to the edge bins. This histogram is taken as the $P$ probability distribution of Eq.~\eqref{eq:KL_definition}. For $Q$, the expected distribution, we turn to the ensemble of simulations, either FFP7 or
the Gaussian realizations. We calculate the histogram for each simulation (using the same range and binning), and take $Q$ to be the mean of all histograms. These histograms are shown in Fig.~\ref{subfig:maps_histograms} together with error bars showing the 5--95\% range for the FFP7 set.
It appears that the histogram of the SMICA map deviates from the reference histograms by~$\approx2\sigma$ primarily in the vicinity of the peak of the distribution. However, this estimation relies on a local feature. The KL divergence provides us with a recipe to sum all the deviations from the entire range of the distribution with appropriate weights.

Before using Eq.~\eqref{eq:KL_definition} to calculate the KL divergence, it is necessary to pay particular attention to bins in which either $P$ or $Q$ has small values.
The case in which $P_i=0$ is not problematic since $P_i\log P_i \to 0$ in this limit.  Bins for which $Q_i=0$, however, should not be included since the
KL divergence is logarithmically divergent as $Q_i \to 0$.  Such a result is not unreasonable since it is impossible to draw to a bin if its probability is
strictly $0$.  In practice, however, small values of $Q_i$ are merely a consequence of the size of our ensemble.  We have chosen to ignore bins for
which $Q_i<5$~pixels in order to minimize the sensitivity to small non-statistical fluctuations in the extreme tails of the $Q$ distribution.

\subsection{The Basic Results}
\label{subsec:The_Basic_Results}

We have calculated the KL divergences between the SMICA histogram and the histograms made from the FFP7 and the Gaussian ensembles. After normalization using the number of valid pixels, $N$, and the number of bins, $N_b$, in Eq.~\eqref{eq:alpha_definition}, we have obtained $\alpha=6.47$ and $6.28$, respectively. If these values were significantly larger than the expected order of magnitude, we would conclude that the distributions were in disagreement. Since this is not the case, we must compare the results to ensembles of values of $\alpha$.
In order to calculate the $p$-values, we repeat the calculation of the KL divergence, replacing the distribution of the map, $P$, with that
of each of the random simulations. This results in two histograms of normalized $K$ values, i.e.\ $\alpha$ values, for FFP7 (Gaussian) maps compared to the FFP7 (Gaussian) mean distribution,
shown in Fig.~\ref{subfig:KL_histograms}. It is evident that, as expected, $\alpha\lesssim10$ for most simulations. We find that $5.6\%$ of the FFP7 simulations and $6.3\%$ of the Gaussian simulations get a higher KL
divergence than the SMICA map. We see that the KL divergence of the SMICA map from the expected distribution is not significant. As expected, differences
between the two reference ensembles, the FFP7 simulations and the pure Gaussian realizations, are quite small.
In order to demonstrate explicitly the similarity between the two ensembles with respect to the KL divergence, we have also tested each of the FFP7 simulations
against the mean distribution of the Gaussian ensemble and vice versa. The results of this calculation are extremely similar to those shown in Fig.~\ref{subfig:KL_histograms}, and we can conclude that the added complexity of the FFP7 simulations relative to that of simple Gaussian realizations plays a minor role at this resolution.

\begin{figure}
	\centering
	\includegraphics[width=\halfW]{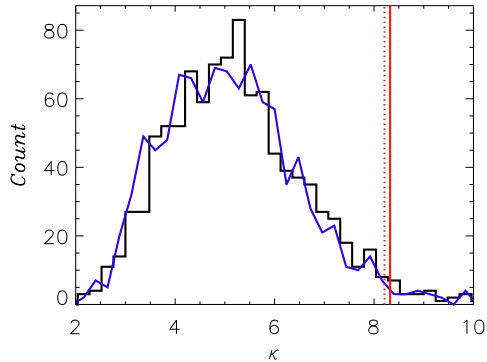}
	\caption{Histograms of KS test values for the FFP7 ensemble (\emph{black}) and the Gaussian ensemble (\emph{blue}). The values for the SMICA map are shown as red vertical
lines, compared to the FFP7 mean distribution (\emph{solid}) and to the Gaussian mean distribution (\emph{dotted}).}
	\label{fig:KSResults}
\end{figure}

In addition to the KL divergence, and as a basis for comparison, we also use the KS test to compare between the histogram of the SMICA map and the mean histogram for each of the ensembles. The KS test, defined in Eq.~\eqref{eq:KS_definition}, requires the use of the CDF. For the SMICA map, the CDF is calculated from the data without any binning. The reference CDF, however, is calculated by first fitting the mean histogram of the ensemble (either FFP7 or the Gaussian realizations) to a Gaussian, and then using the fitted parameters in the expression for a Gaussian CDF. As in the case of the KL divergence, for each ensemble, we compare the SMICA map to the mean histogram of the ensemble and also create a histogram of KS test values by taking each realization separately and comparing it to the mean. The results are shown in Fig.~\ref{fig:KSResults}.
The KS test values we get when comparing the SMICA map to the FFP7 and Gaussian simulations are $\kappa=8.32$ and $8.21$, respectively. The corresponding $p$-values are $3.0\%$ and $2.6\%$. Again we see that the results for the two ensembles are in good agreement. Moreover, while the $p$-values of the KS test are lower than those of the KL divergence, the SMICA map still appears to be consistent with the reference ensembles and not anomalous.

\begin{figure}
	\centering
	\subfigure[]{
		\includegraphics[width=\thirdW,angle=90]{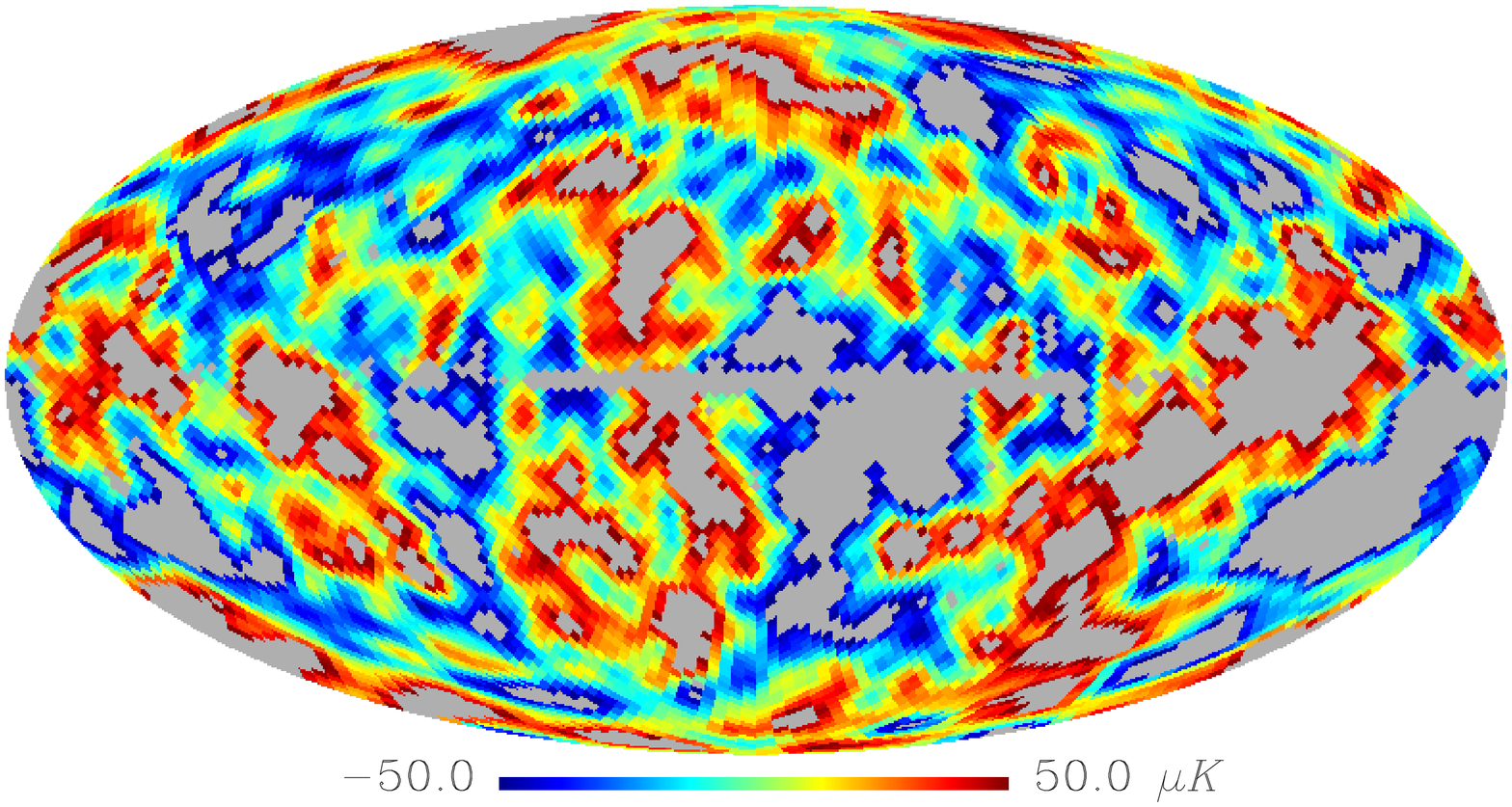}
		\label{subfig:smica_-50_50}
	}
	\subfigure[]{
		\includegraphics[width=\thirdW,angle=90]{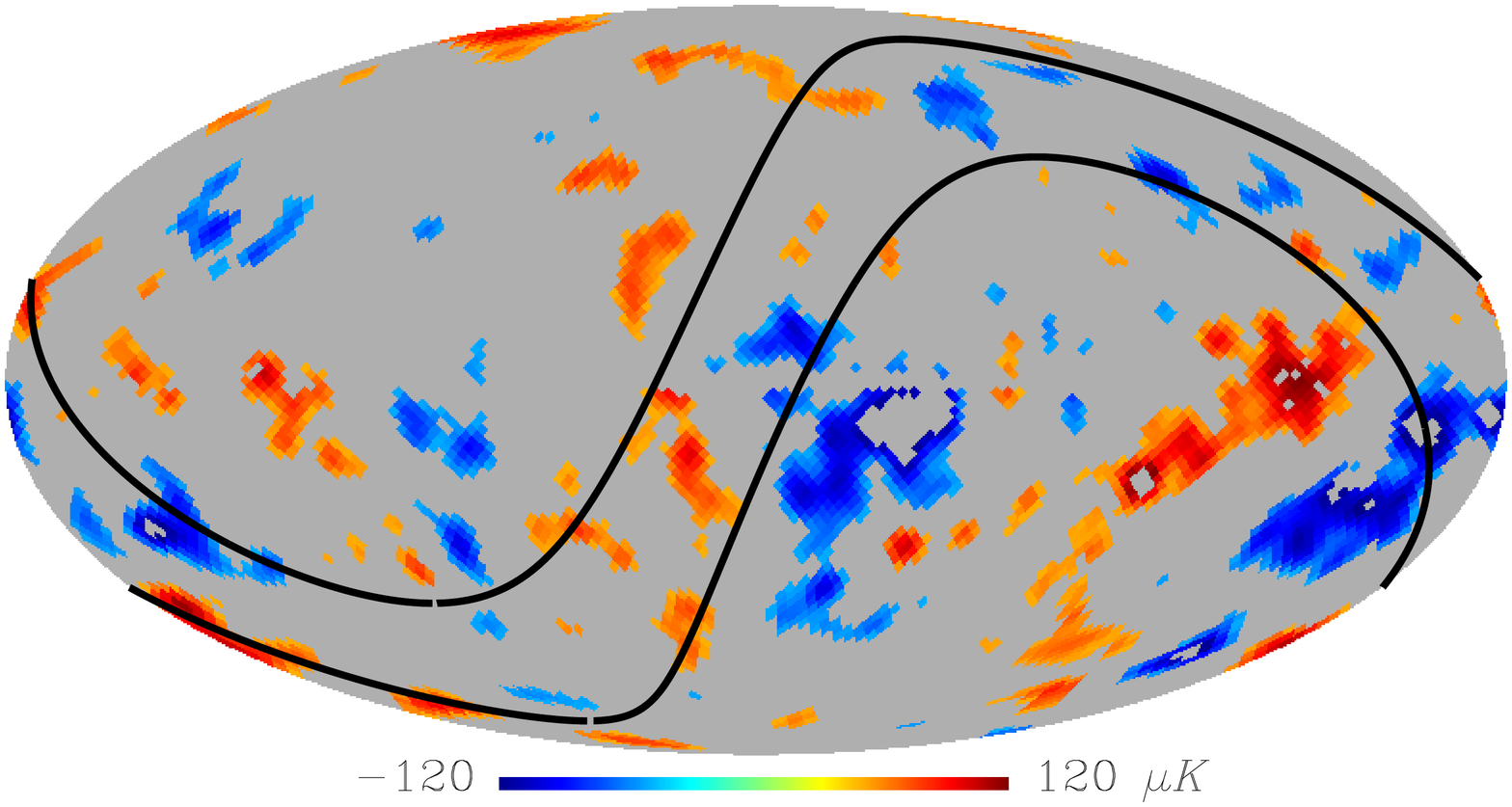}
		\label{subfig:smica_abs_50_120}
	}
	\caption{The SMICA map, showing only those regions where \subref{subfig:smica_-50_50}~$|T|\le50\uK$ and \subref{subfig:smica_abs_50_120}~$50\uK\le|T|\le120\uK$. The small Galactic mask used in the analysis appears as a thin horizontal gray line in the center of the maps, and the masked pixels are not included in any of the temperature ranges. In panel~\subref{subfig:smica_abs_50_120}, the black curves mark the location of the ecliptic plane.}
	\label{fig:smica_regions}
\end{figure}

As we can see from Fig.~\ref{subfig:maps_histograms}, there are well-defined temperature ranges in which the SMICA histogram is above or
below the reference.  Thus, in Fig.~\ref{fig:smica_regions} we plot the SMICA map, showing only the temperature range $|T|\le50\uK$ where the SMICA
histogram is above the reference and the temperature range $50\uK\le|T|\le120\uK$ where it is below.
We see that there is no apparent tendency for the contributions from either of these temperature ranges to be localized in specific regions of the sky. We do, however, pay special attention to the region of the ecliptic plane. As is apparent from fig.~3 of~\cite{Planck15Maps}, the SMICA map is susceptible to contamination from foreground residuals in the region of the ecliptic. We therefore include in Fig.~\ref{subfig:smica_abs_50_120} curves showing the location of the ecliptic plane and suggest that the number of cold spots in the ecliptic band might be unexpected. As it is not the focus of this work, we have not performed any quantitative analysis regarding the spatial distribution of hot or cold regions of the map. The maps in Fig.~\ref{fig:smica_regions} provide
an additional general indication that the SMICA temperature map is not anomalous. Nevertheless, we again emphasize that small foreground residuals, like those suspected to lie in the area of the ecliptic plane, while insignificant to the analysis of temperature fluctuations, can become extremely important when analyzing the CMB polarization pattern, specifically $B$-mode polarization.

\subsection{Interchangeability of $P$ and $Q$}
\label{sub:interchangeability_of_P_and_Q}

\begin{figure}
	\centering
	\subfigure[]{
		\includegraphics[width=\halfW]{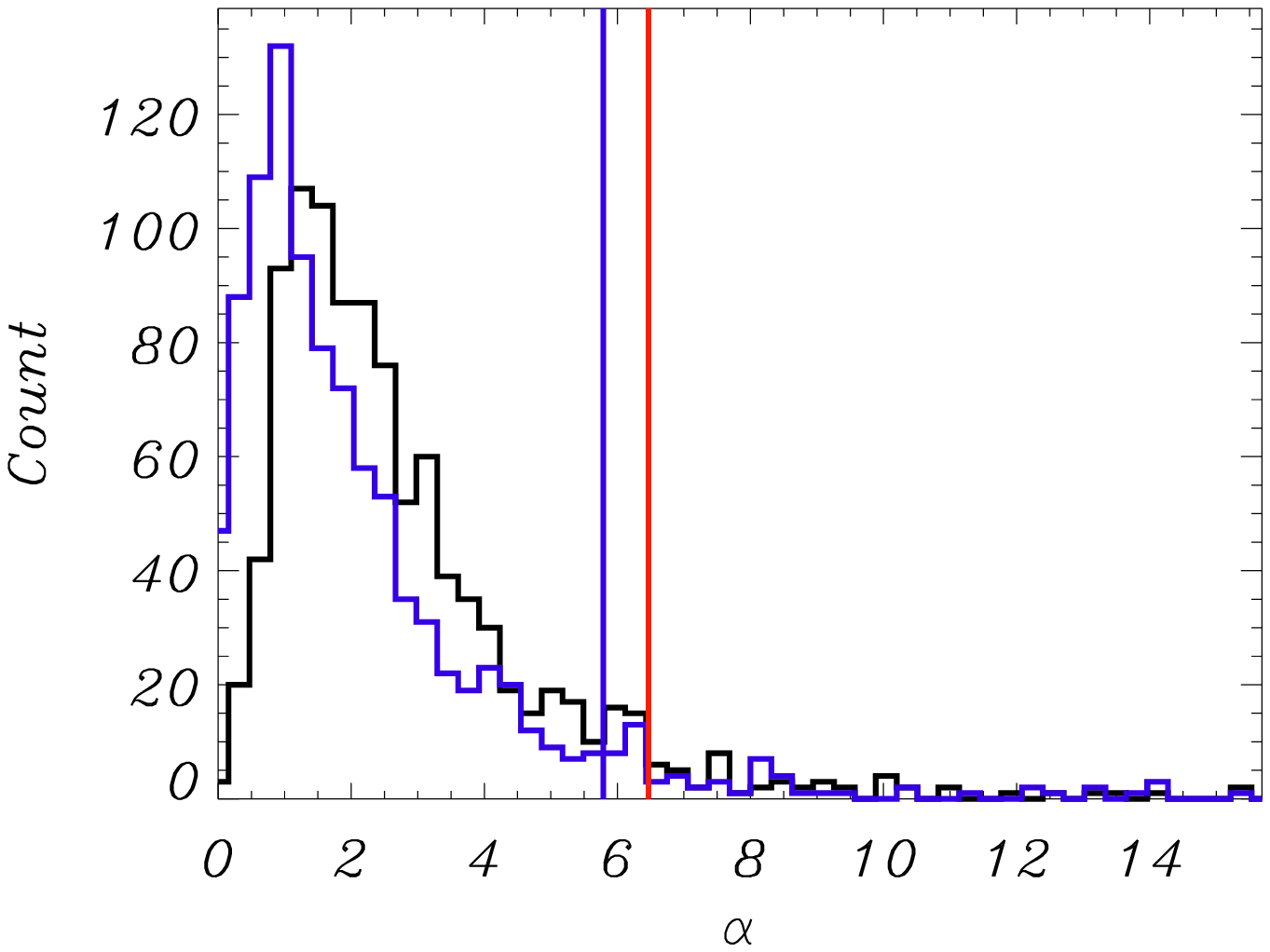}
		\label{subfig:PQ_and_QP}
	}
	\subfigure[]{
		\includegraphics[width=\halfW]{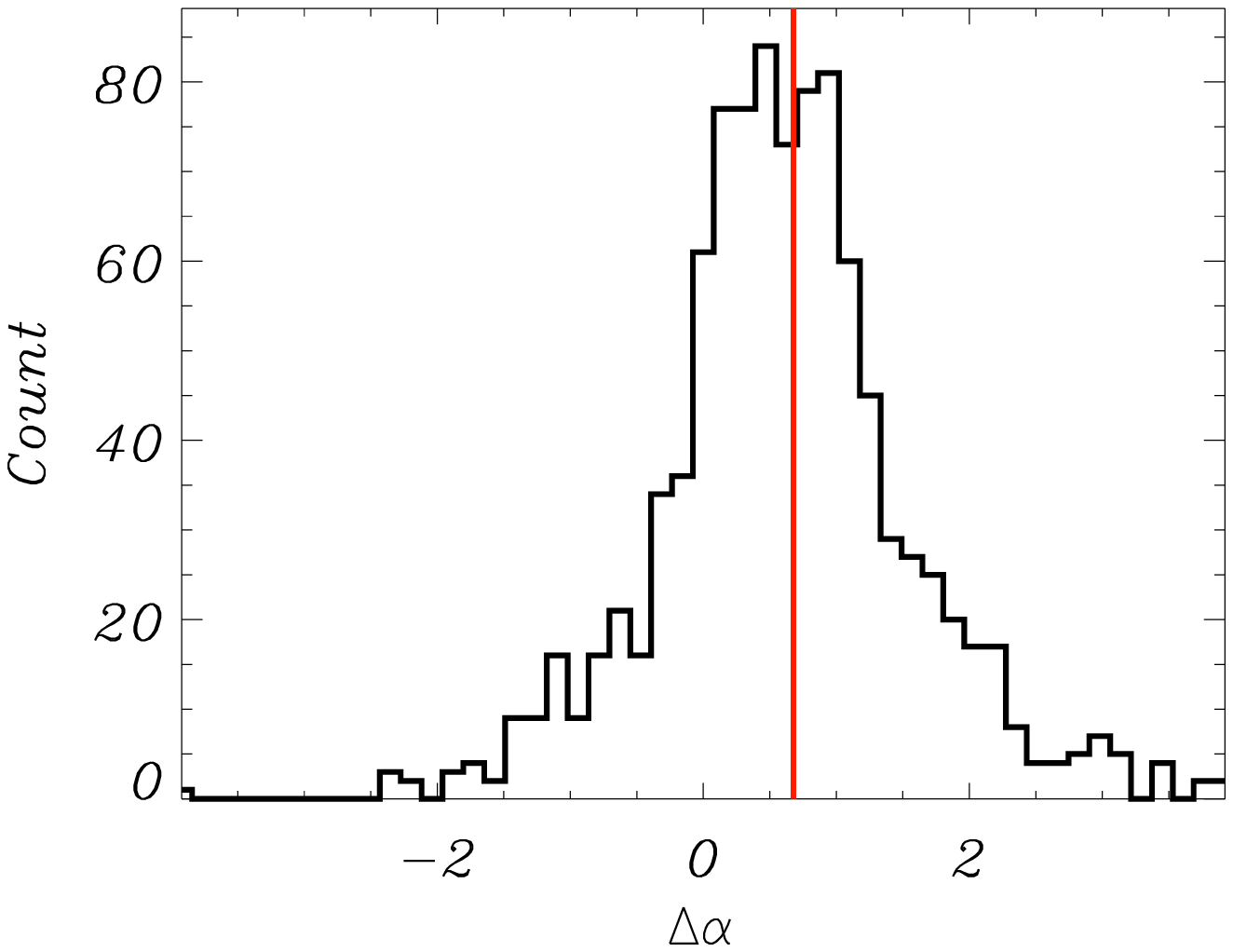}
		\label{subfig:PQ_-_QP}
	}
	\caption{\subref{subfig:PQ_and_QP}~Histograms for the usual KL divergence $K(P\|Q)$ (\emph{black}) and for the reversed divergence $K(Q\|P)$
	(\emph{blue}). The red and blue vertical lines are the values for the SMICA map for the normal and reversed tests, respectively. All $K$ values have been normalized using Eq.~\eqref{eq:alpha_definition}.
\subref{subfig:PQ_-_QP}
	~Histogram of the normalized difference $\alpha(P\|Q)-\alpha(Q\|P)$. The vertical line is the value for the SMICA map.}
	\label{fig:PQQP}
\end{figure}

As has been noted above, the KL divergence is not symmetric with respect to the interchange of the distributions $P$ and $Q$ except in the limit $P \to Q$.
This is a reminder of the fact that the KL divergence is not a true metric of the distance between $P$ and $Q$.   So far, we have followed the common practice of taking $P$ to be the distribution of the data and $Q$ the expected distribution~\cite{kl}.  However, it is worth checking what happens when these roles are reversed.  We have performed two tests involving interchange of the two distributions. First, we simply calculate the KL divergence $K(Q\|P)$, where $P$ is again the SMICA histogram and $Q$ is the mean histogram of the FFP7 ensemble.\footnote{Note that with the roles reversed, bins with $P_i < 5$ are now ignored and bins with small values of $Q_i$ are all counted.} This value is then compared to the ensemble of values computed with $P$ replaced by each of the FFP7 maps. The
resulting histogram, after normalization using Eq.~\eqref{eq:alpha_definition}, is presented in Fig.~\ref{subfig:PQ_and_QP} together with the histogram of $K(P\|Q)$ (presented above) as reference.
We can see that the two histograms and the corresponding $p$-values are similar. The value $p = 6.5\%$ was obtained for the reversed test; the
$p$-value for the normal test is $5.6\%$ as stated above.  It is apparent that, although similar, the reversed histogram is slightly but consistently shifted towards smaller $\alpha$ values than the normal histogram. The value for the SMICA map is also lower for the reversed test.  However, the SMICA is a single map, and the shift between histograms only indicates a statistical shift for the whole ensemble. Therefore, the second test is to examine the difference $\Delta\alpha = \alpha(P\|Q)-\alpha(Q\|P)$ between the normal and reversed normalized KL divergences of the \emph{same} map. Fig.~\ref{subfig:PQ_-_QP} shows the resulting histogram for the FFP7 ensemble together with the value for SMICA. The SMICA map shows a highly standard $\Delta\alpha$, yielding a $p$-value of $49.4\%$. A similar test performed versus the Gaussian ensemble gives very similar results.

The tendency of the KL divergences to become smaller when $P$ and $Q$ are interchanged can be understood easily. Since $P$ is calculated from a single map, it tends to fluctuate more than $Q$, which is the mean of all ensemble distributions. Therefore, when $P$ appears only inside the logarithm, as is the case in the reversed $K(Q\|P)$, the fluctuations are suppressed relative to the normal $K(P\|Q)$. We see here that the SMICA map not only shows the expected qualitative behavior upon interchanging $P$ and $Q$, it is also quantitatively shifted by the expected amount. While the KL divergence in general is not symmetric under the interchange of $P$ and $Q$, we conclude that when testing the one-dimensional temperature distribution of the CMB on large scales, reversing the two makes little difference.

\section{Discussion} % (fold)
\label{sec:discussion}

We have discussed the applicability of the Kullback--Leibler divergence for the assessment of departures form Gaussianity of CMB temperature
maps on large scales.  We have illustrated this on the SMICA map, comparing it to both the set of FFP7 simulations and a set of $10^3$ Gaussian draws.
We have shown that it is consistent with each of these reference sets to a level of about 6\%.  We have used the KL divergence to compare the FFP7 and Gaussian reference sets and have shown that they are in good agreement.  This suggests that the additional instrumental effects and foreground residuals included in the FFP7 simulations are unimportant on the scales considered here.  Since the KL divergence is not symmetric in $P$ and $Q$, we have performed tests to demonstrate that their interchange has little effect on these conclusions.  Finally, we have repeated these calculations using the Kolmogorov--Smirnov test.  The resulting
$p$-value of about 3\% suggests that the differences between the two tests are not large.  We note that there is no guarantee that these tests will always
give similar results.  For example, the KL divergence is likely to be far more sensitive than the KS test for situations where there are large relative differences in the small amplitude tails of the distributions.  We have also repeated all the tests on the CMB data of the 2015 release from Planck, which recently became publicly available.\footnote{See the Planck Legacy Archive \url{http://pla.esac.esa.int/pla/}.} The results on the 2015 data set are in very good agreement with those reported here.

The difficulty in devising tests for the assessment of non-Gaussianity of the temperature and polarization maps of the CMB lies in our ignorance of
the nature of the non-Gaussian residuals from foregrounds and systematic effects that could propagate to the maps.  In such circumstances, it seems advisable to adopt a procedure that uses as much information in the maps as possible.  With its connection to the intrinsic and cross-entropy of the distributions $P$ and $Q$,
the KL divergence would appear to be the natural choice.  Given the correlations between the pixels of the CMB, a consequence of a random draw in harmonic space, this is not necessarily the case.  However, the non-locality of the KL divergence and its sensitivity to the tails of the distributions still suggest that it is a valuable complement to the KS test and might be a useful alternative.  Indeed, one should utilize a variety of methods and tests to identify possible contamination of the cosmological product.  Obviously, \emph{any\/} suggestion of an anomalous result would indicate the need for more sophisticated analyses to assess the quality of the CMB maps.

\acknowledgments

We would like to thank P. Naselsky for his enthusiastic support and encouragement of the work reported in this paper and for many valuable scientific discussions.

This work is based on observations obtained with Planck,\footnote{\url{http://www.esa.int/Planck}} an ESA
science mission with instruments and contributions directly funded by ESA
Member States, NASA, and Canada. The development of Planck has been
supported by: ESA; CNES and CNRS / INSU-IN2P3-INP (France); ASI, CNR, and
INAF (Italy); NASA and DoE (USA); STFC and UKSA (UK); CSIC, MICINN and JA
(Spain); Tekes, AoF and CSC (Finland); DLR and MPG (Germany); CSA (Canada);
DTU Space (Denmark); SER/SSO (Switzerland); RCN (Norway); SFI (Ireland);
FCT/MCTES (Portugal); and PRACE (EU). A description of the Planck Collaboration and a list of its members,
including the technical or scientific activities in which they have been
involved, can be found at the Planck web page.\footnote{\url{http://www.cosmos.esa.int/web/planck/planck-collaboration}}

We acknowledge the use of the NASA Legacy Archive for Microwave Background Data Analysis (LAMBDA). Our data analysis made use of the GLESP package\footnote{\url{http://www.glesp.nbi.dk/}}~\cite{Glesp}, and of \textsc{HEALPix}~\cite{HEALPix}. This work is supported in part by Danmarks Grundforskningsfond which allowed the establishment of the Danish Discovery Center, FNU grants 272-06-0417, 272-07-0528 and 21-04-0355, the National Natural Science Foundation of China (Grant No.\ 11033003), the National Natural Science Foundation for Young Scientists of China (Grant No.\ 11203024) and the Youth Innovation Promotion Association, CAS.


\begin{thebibliography}{27}

\bibitem{Planck1}
{Planck Collaboration}, P.~A.~R. {Ade} et~al., {\em \aap} {\bf 571} (2014) A1
  [\href{http://arxiv.org/abs/1303.5062}{{\tt arXiv:1303.5062}}].

\bibitem{Planck2}
{Planck Collaboration}, P.~A.~R. {Ade} et~al., {\em \aap} {\bf 571}
  (2014) A24 [\href{http://arxiv.org/abs/1303.5084}{{\tt
  arXiv:1303.5084}}].

\bibitem{Planck_is}
{Planck Collaboration}, P.~A.~R. {Ade} et~al., {\em \aap} {\bf 571} (2014) A23 [\href{http://arxiv.org/abs/1303.5083}{{\tt arXiv:1303.5083}}].

\bibitem{Planck_is2015}
{Planck Collaboration}, P.~A.~R.~{Ade} et~al.\ [\href{http://arxiv.org/abs/1506.07135}{{\tt arXiv:1506.07135}}].

\bibitem{low_quadrupole}
G.~{Efstathiou}, {\em \mnras} {\bf 346} (2003) L26--L30
  [\href{http://arxiv.org/abs/astro-ph/0306431}{{\tt astro-ph/0306431}}].

\bibitem{wmap9b}
C.~L. {Bennett} et~al., {\em \apjs} {\bf 208} (2013)
  20 [\href{http://arxiv.org/abs/1212.5225}{{\tt arXiv:1212.5225}}].

\bibitem{Copi1}
C.~J. {Copi}, D.~{Huterer}, and G.~D. {Starkman},  {\em \prd} {\bf 70} (2004)
  043515 [\href{http://arxiv.org/abs/astro-ph/0310511}{{\tt
  astro-ph/0310511}}].

\bibitem{mhansen}
P.~{Naselsky}, M.~{Hansen}, and J.~{Kim},  {\em
  \jcap} {\bf 9} (2011) 12 [\href{http://arxiv.org/abs/1105.4426}{{\tt
  arXiv:1105.4426}}].

\bibitem{Eriksen2003} 
  H.~K.~{Eriksen} et~al.,
	{\em \apj} {\bf 605} (2004) 14--20 
  [{\em \apj} {\bf 609} (2004) 1198--1199]
  [\href{http://arxiv.org/abs/astro-ph/0307507}{{\tt astro-ph/0307507}}].

\bibitem{Hansen2004} 
  F.~K.~Hansen, A.~J.~Banday and K.~M.~G\'orski,
  {\em \mnras} {\bf 354} (2004) 641--665
  [\href{http://arxiv.org/abs/astro-ph/0404206}{{\tt astro-ph/0404206}}].

\bibitem{WMAP7:powerspectra}
D.~{Larson} et~al., {\em \apjs} {\bf 192} (2011) 16 [\href{http://arxiv.org/abs/1001.4635}{{\tt arXiv:1001.4635}}].

\bibitem{Pref.Direction1}
K.~{Land} and J.~{Magueijo}, {\em \prl} {\bf 95} (2005)
  071301 [\href{http://arxiv.org/abs/astro-ph/0502237}{{\tt
  astro-ph/0502237}}].

\bibitem{Pref.Direction2}
K.~{Land} and J.~{Magueijo}, {\em \mnras}
  {\bf 378} (2007) 153--158
  [\href{http://arxiv.org/abs/astro-ph/0611518}{{\tt astro-ph/0611518}}].

\bibitem{Copi2}
C.~J. {Copi} et~al., {\em Advances in Astronomy} {\bf 2010}
  (2010) 92 [\href{http://arxiv.org/abs/1004.5602}{{\tt arXiv:1004.5602}}].

\bibitem{Akrami2014} 
  Y.~Akrami et~al.,
  {\em \apjl} {\bf 784} (2014) L42 
  [\href{http://arxiv.org/abs/1402.0870}{{\tt arXiv:1402.0870}}].

\bibitem{parity1}
J.~{Kim} and P.~{Naselsky},
  {\em \prd} {\bf 82} (2010) 063002
  [\href{http://arxiv.org/abs/1002.0148}{{\tt arXiv:1002.0148}}].

\bibitem{parity2}
J.~{Kim} and P.~{Naselsky}, {\em
  \apjl} {\bf 714} (2010) L265--L267
  [\href{http://arxiv.org/abs/1001.4613}{{\tt arXiv:1001.4613}}].

\bibitem{vielva2003}
P.~{Vielva} et~al.,  {\em
  \apj} {\bf 609} (2004) 22--34
  [\href{http://arxiv.org/abs/astro-ph/0310273}{{\tt astro-ph/0310273}}].

\bibitem{Coldspot1}
M.~{Cruz} et~al., {\em \apj} {\bf 655} (2007) 11--20
  [\href{http://arxiv.org/abs/astro-ph/0603859}{{\tt astro-ph/0603859}}].

\bibitem{Planck_bicep}
{Planck Collaboration}, R.~{Adam} et~al.\ [\href{http://arxiv.org/abs/1409.5738}{{\tt arXiv:1409.5738}}].

\bibitem{inflation}
A.~H. {Guth}, {\em \prd} {\bf 23} (1981) 347--356; \\
A.~D. {Linde}, {\em Physics Letters B} {\bf 108} (1982) 389--393; \\
A.~D. {Linde}, {\em Physics Letters B} {\bf 129} (1983) 177--181; \\
A.~{Albrecht} and P.~J. {Steinhardt}, {\em \prl} {\bf 48} (1982) 1220--1223; \\
V.~F. {Mukhanov} and G.~V. {Chibisov}, {\em J.\ Exp.\ Theor.\ Phys.\ Lett.} {\bf 33} (1981) 532; \\
A.~H. {Guth} and S.-Y. {Pi}, {\em \prl} {\bf 49} (1982) 1110--1113; \\
J.~M. {Bardeen}, P.~J. {Steinhardt}, and M.~S. {Turner}, {\em \prd} {\bf 28} (1983) 679--693; \\
A.~{Linde} and V.~{Mukhanov}, {\em \prd} {\bf 56} (1997) 535
  [\href{http://arxiv.org/abs/astro-ph/9610219}{{\tt astro-ph/9610219}}].

\bibitem{bicep}
{BICEP2 Collaboration}, P.~A.~R. {Ade} et~al., {\em \prl} {\bf 112} (2014) 241101
  [\href{http://arxiv.org/abs/1403.3985}{{\tt arXiv:1403.3985}}].

\bibitem{Planck_cp}
{Planck Collaboration}, P.~A.~R. {Ade} et~al., {\em \aap} {\bf 571} (2014) A16
  [\href{http://arxiv.org/abs/1303.5076}{{\tt arXiv:1303.5076}}].

\bibitem{kl}
S.~Kullback and R.~A. Leibler, {\em
  Ann.\ Math.\ Statist.} {\bf 22} (1951) 79--86.

\bibitem{HEALPix}
K.~M. {G{\'o}rski} et~al., {\em \apj} {\bf 622} (2005) 759--771
  [\href{http://arxiv.org/abs/astro-ph/0409513}{{\tt astro-ph/0409513}}].

\bibitem{Planck15Maps}
{Planck Collaboration}, R.~{Adam} et~al.\ [\href{http://arxiv.org/abs/1502.05956}{{\tt arXiv:1502.05956}}].

\bibitem{Glesp}
A.~G. {Doroshkevich} et~al., {\em
  \ijmpd} {\bf 14} (2005) 275--290
  [\href{http://arxiv.org/abs/astro-ph/0305537}{{\tt astro-ph/0305537}}].

\end{thebibliography}
\end{document}